\documentclass[12pt]{article}
\usepackage{amsmath}
\usepackage{graphicx}
\usepackage{psfrag}
\usepackage{amssymb}
\newcommand{\rf}[1]{(\ref{#1})}
\newcommand{\bea}{\begin{eqnarray}}
\newcommand{\eea}{\end{eqnarray}}

\newcommand{\e}{\mbox{e}}
\renewcommand{\d}{\mbox{d}}

\renewcommand{\l}{\lambda}
\renewcommand{\L}{\Lambda}
\renewcommand{\b}{\beta}
\renewcommand{\a}{\alpha}

\newcommand{\m}{\mu}

\newcommand{\om}{\omega}
\newcommand{\del}{\delta}
\newcommand{\Del}{\Delta}

\newcommand{\oh}{\frac{1}{2}}

\newcommand{\tr}{\mathrm{tr}\,}
\newcommand{\ra}{\rangle}
\newcommand{\la}{\langle}
\newcommand{\prt}{\partial}
\newcommand{\mi}{\!-\!}
\newcommand{\equ}{\!=\!}
\newcommand{\pl}{\!+\!}
\newcommand{\cuh}{{\{ h \}}}

\newcommand{\cN}{{\cal N}}

\newcommand{\cG}{{\cal G}}

\newcommand{\trho}{{\tilde{\rho}}}

\newcommand{\hT}{{\hat{T}}}
\newcommand{\hH}{{\hat{H}}}

\newcommand{\no}{\nonumber}

\newcommand{\Pint}{{\mathbf{-}}\!\!\!\!\!\!\int}
\newcommand{\PintI}{\Pint_{I_0}}
\newcommand{\dt}{\frac{dt}{\pi i}\,}
\renewcommand{\dh}{\frac{dh}{\pi i}\,}

\newcommand{\Drhro}{\Del {\rho}_{r}}
\newcommand{\fro}{f_{r}}

\def\void{}
\def\labelmark{}

\newenvironment{formula}[1]{\def\labelname{#1}
\ifx\void\labelname\def\junk{\begin{displaymath}}
\else\def\junk{\begin{equation}\label{\labelname}}\fi\junk}%
{\ifx\void\labelname\def\junk{\end{displaymath}}
\else\def\junk{\end{equation}}\fi\junk\labelmark\def\labelname{}}

{\ifx\void\labelname\def\junk{\end{array}\end{displaymath}}
\else\def\junk{\end{array}\right.\end{equation}}
\fi\junk\labelmark\def\labelname{}\def\junk{}
}

\newcommand{\beq}{\begin{formula}}
\newcommand{\eeq}{\end{formula}}
\newcommand{\beqv}{\begin{formula}{}}

\begin{document}

\hfill SPIN-2003/27

\hfill ITP-UU-03/44

\vspace{30pt}

\begin{center}
{\large \bf 3D Lorentzian Quantum Gravity from 
the \\asymmetric ABAB matrix model
\footnote{Presented by J.J. at the Workshop on Random Geometry, 
May 15-17, 2003, Krakow, Poland}%
}
\vspace{30pt}

{\sl  Jan Ambj\o rn$^{1,4}$, Jerzy Jurkiewicz$^2$, Renate Loll$^{3,4}$
and Graziano Vernizzi$^3$}\\

\vspace{24pt}

{\footnotesize
$^1$The Niels Bohr Institute, \\
Blegdamsvej 17, DK-2100 Copenhagen \O , 
Denmark\\ 
{\tt email:}  {\tt ambjorn@nbi.dk}\\
\vspace{3mm}
$^2$Inst. of Physics, Jagellonian University,\\
ul. Reymonta 4, 30-059 Krakow, Poland\\
{\tt email:} {\tt jurkiewicz@th.if.uj.edu.pl}\\
\vspace{3mm} 
$^3$Institute for Theoretical Physics, Utrecht University,\\
Leuvenlaan 4, NL-3584 CE Utrecht, The Netherlands\\ 
{\tt email:} {\tt r.loll@phys.uu.nl, g.vernizzi@phys.uu.nl}\\
\vspace{3mm}
$^4$Perimeter Institute for Theoretical Physics,\\
35 King Street North, Waterloo, ON, Canada N2J 2W9\\
{\tt email:}{\tt jambjorn@perimeterinstitute.ca, \\
~~~rloll@perimeterinstitute.ca}
}
\end{center}

\vspace{48pt}

\begin{center}
{\bf Abstract}
\end{center}

The asymmetric ABAB-matrix model describes the transfer matrix
of three-dimensional Lorentzian quantum gravity. 
We study perturbatively the scaling
of the ABAB-matrix model in the neighbourhood of its
symmetric solution and deduce the associated renormalization 
of three-dimensional Lorentzian quantum gravity.
 
\vspace{18pt}
\noindent
pacs: 04.60Gw, 04.20Gz, 04.60Kz, 04.60Nc

\newpage
  
\section{Introduction}
Matrix models have been very useful in the study of 
the quantum geometry of two-dimensional quantum 
gravity. In \cite{ajlv} this program was 
extended to three-dimensional quantum gravity.
It was shown how the so-called ABAB two-matrix model
describes the transfer matrix of three-dimensional 
quantum gravity.\footnote{Previous work on 3D quantum
gravity in terms of Lorentzian triangulations can
be found in \cite{ajl2,ajl2.5}.}
More precisely, a non-perturbative, 
background-independent definition of quantum gravity,
which emphasizes the causal structure of space-time
and which allows rotations between Lorentzian and 
Euclidean signature, was proposed in \cite{ajl1,ajl1.5},
generalizing an explicitly solvable two-dimensional model 
with these features \cite{al}. In the model, which  has an 
$UV$ lattice cut-off $a$ which should be taken to zero in the 
continuum limit, one can define 
the concept of proper time. In the Euclidean sector 
the corresponding evolution operator is defined in terms of 
the transfer matrix $\hT$ describing the transition between quantum 
states at (proper) time $n \cdot a$ and (proper) time $(n\pl 1)\cdot a$.
The transfer matrix  is related to the quantum Hamiltonian of the system by 
\beq{0.1}
\hT = \e^{-a \hH}.
\eeq
The ABAB model is defined by the two-matrix integral 
\bea\label{I.1}
Z(\a_1,\a_2,\b) &=& \e^{-M^2 F(\a_1,\a_2,\b)} \nonumber \\
&=&\int \d A \d B \;
\e^{-M \tr (\oh (A^2+B^2) -\frac{\a_1}{4}A^4-\frac{\a_2}{4}B^4 -
\frac{\b}{2} ABAB)}.
\eea
Under the assumption discussed in \cite{ajlv} 
the free energy $F(\a_1,\a_2,\b)$ is related to the matrix 
elements of the transfer matrix $\hT$ in a way reviewed in 
the next section. The matrix model \rf{I.1} has a scaling 
limit for $\a_1 \equ \a_2$ which was  analyzed in \cite{kz}.
This allowed us in \cite{ajlv} to determine the corresponding 
phase diagram for the three-dimensional quantum gravity model
and to map the bare coupling constants of the gravity 
model to the matrix model coupling constants $\a_1\equ \a_2$ and 
$\b$ \cite{newpaper}. However, in order to study details of the scaling 
relevant to three-dimensional quantum gravity we have 
to study the matrix model  for $\a_1 \neq \a_2$. In the scaling 
limit of interest for us both $\a_1$ and $\a_2$ will scale to 
 a critical value $\a_c$, but independently. Since we are interested
only in the behaviour of the theory near the symmetric
solution we need only the perturbative expansion around this solution
rather than the complete solution in the asymmetric case\footnote{While 
writing this article the asymmetric ABAB matrix model
has been solved by Paul Zinn-Justin \cite{newPZJ}. The behaviour 
close to the symmetric line $\a_1=\a_2$ is the same as the one
reported here and to extract it one has to expand the elliptic 
functions which appear in the solution, an effort comparable to 
the one used here.}. 

The rest of this 
article is organized as follows. In sec.\ \ref{model}
we review shortly the non-perturbative definition of 
three-dimensional Lorentzian quantum gravity \cite{ajl1,ajl2}
and its relation to the ABAB matrix model. In sec.\ \ref{ABAB}
we review the machinery needed to solve the ABAB matrix model
for a symmetric choice of coupling constants \cite{kz}.
In Sec.\ \ref{asym} we discuss the solution of the 
general ABAB matrix model, and in Sec.\ \ref{expander} we 
expand around the symmetric critical point relevant for 
three-dimensional quantum gravity. In Sect.\ \ref{discuss}
we discuss how to extract information about 
the transfer matrix of 3D gravity, knowing the free energy of the 
asymmetric ABAB matrix model.

\section{Quantum gravity and the ABAB matrix model}\label{model}

Simplicial Lorentzian quantum gravity in three dimensions
is defined in the following 
way: the spatial hypersurfaces of constant proper-time are 
two-dimensional equilateral triangulations. Such triangulations 
define uniquely a two-dimensional geometry. It is known that 
this class of geometries describes correctly the quantum aspects 
of two-dimensional Euclidean gravity. It is also known that the 
description of two-dimensional Euclidean quantum gravity in terms
of the class of (generalized) triangulations is quite robust.
In \cite{ajl1} we used this universality in the 
following way: the two-dimensional geometry of the spatial hypersurfaces 
is represented by quadrangulations and it was shown that 
it is possible to connect any such pair of quadrangulations 
by a set of three-dimensional generalized ``simplices''. 
More precisely, let $a$ be the lattice
spacing separating two neighbouring spatial hypersurfaces at 
(proper)-times 
$t$ and $t+a$. Then each square at $t$ is connected to a vertex at
$t+a$ and each square at $t+a$ is connected 
to a vertex at proper-time $t$, forming pyramids and inverted 
pyramids. A further needed three-dimensional 
building block is a tetrahedron connecting a spatial link at $t$ to 
a spatial link at $t\pl a$. The proper-time 
propagator for (regularized) three-dimensional quantum gravity 
between two spatial hypersurfaces separated by a proper time $T\equ 
n \cdot a$ is obtained by inserting $n\mi 1$ intermediate 
spatial hypersurfaces and summing over all possible geometries 
constructed as described above. The weight of each geometry is
given by the Einstein action, here conveniently the Regge action 
for piecewise linear geometries. According to \cite{ajl1.5,ajl2}, 
the contribution to
the action from a single discrete time step is given by
\beq{act}
S=-\kappa (N_t +N_{t+a}-N_{22})+\lambda (N_t+N_{t+a}+\frac{1}{2}
N_{22}),
\eeq
where $N_t$, $N_{t+a}$ and $N_{22}$ denote the number of pyramids,
upside-down pyramids and of tetrahedra contained in the slice
$[t,t+a]$, and $\kappa$ and $\lambda$ are the dimensionless 
bare inverse
gravitational and bare cosmological constant in three
space-time dimensions.
The naive continuum limit is obtained 
by scaling the lattice spacing $a \to 0$ while keeping $T \equ n\cdot a$
fixed. However, different scaling relations between $T$ and $a$ 
might in principle be possible\footnote{In two-dimensional {\it Euclidean}
quantum gravity
the proper-time $T$ scales anomalously and one has to keep $n\sqrt{a}$ fixed.
This is in contrast to the situation in two-dimensional {\it Lorentzian}
quantum gravity as defined in \cite{al} where the proper-time $T$
scales canonically. The relation between the two models is well 
understood \cite{ackl}.}.

Let $g_t$ and $g_{t+a}$ be spatial two-geometries at $t$ and $t\pl a$,
i.e.\ two quadrangulations and let $\la g_{t+a}|\hT |g_t\ra$ be the 
transition amplitude or proper time propagator from $t$ to 
$t\pl a$. By definition, $\hT$ is 
the transfer matrix in the sense of Euclidean lattice theory, and 
it satisfies the axioms of a transfer matrix \cite{ajl2}. In the case where 
the spatial topology is that of $S^2$ it was argued in \cite{ajlv}
that the continuum limit could be obtained 
as the large $N$ scaling limit of the matrix model \rf{I.1}.
Let $N_t$ and $N_{t+a}$ denote the number of squares in the 
quadrangulations associated with $g_t$ and $g_{t+a}$. The 
two-volumes of the corresponding geometries are thus $N_t a^2$ and 
$N_{t+a}a^2$, respectively, and the relation to $F(\a_1,\a_2,\b)$ 
defined by \rf{I.1} is 
\beq{II.1}
F(\a_1,\a_2,\b) = \sum_{g_t, g_{t+a}} \e^{-z_t N_t(g_t)-
z_{t+a}N_{t+a}(g_{t+a})} 
\la g_{t+a} | \hT | g_{t}\ra, 
\eeq
where $z_t$ and $z_{t+a}$ are dimensionless boundary cosmological 
constants.  
The na\"ive relation between the matrix model coupling constants 
and the bare dimensionful
gravitational and cosmological coupling constants $G^{(0)}$ and 
$\L^{(0)}$ of
three-di\-men\-sio\-nal gravity is 
\beq{II.2}
\a_1 = \e^{\kappa-\lambda-z_t},~~~~
\a_2= \e^{\kappa-\lambda-z_{t+a}},~~~~
\b= \e^{-(\oh \lambda+\kappa)},
\eeq
where 
\bea\label{II.3}
\kappa= \frac{a}{4\pi G^{(0)}}\left( -\pi+3\cos^{-1}\frac{1}{3} \right)
,~~~
\lambda = \frac{a^3 \L^{(0)}}{24\sqrt{2}\pi}
,~~~
z_t= a^2 Z_t^{(0)}.
\eea
In this paper we shall discuss the non-perturbative renormalization 
of the coupling constants.
In principle we are interested in the limit $z_t\equ  z_{t+a}\equ 0$,
i.e.\ $\a_1\equ \a_2$. 
However, in order to be able to extract the information about 
the scaling of the boundary
cosmological constants  we have to keep $z_t$ and $z_{t+a}$ different 
from zero at intermediate steps. Thus these 
{\it boundary cosmological constants}
should be viewed as source terms for the boundary area operator.

\section{The symmetric case: {\boldmath{$\a_1= \a_2= \a$}}}\label{ABAB}

Let us for later convenience 
shortly review the technique for solving the matrix model 
\rf{I.1}  used in \cite{kz} (based on earlier results \cite{ksw}).

By a character 
expansion of the term $e^{\oh \b M \tr ABAB}$ one can write
\beq{3.1}
Z(\a_1,\a_2,\b) \sim 
\sum_{\{ h\}} \Big(\frac{M \b}{2}\Big)^{\oh \sum h_i -\frac{M(M-1)}{4}}\; 
c_{\{ h\}}\; 
R_{\{ h\} } (\a_1)R_{\{ h\} } (\a_2),
\eeq
where the sum is over the representations of $GL(M)$, characterized by 
the shifted highest weights $h_i \equ m_i +M-i$, $(i\equ 1,\ldots,M)$, where 
the $m_i$ are the standard highest weights and where the large-$M$ limit of
the coefficient $c_\cuh$ is 
\beq{3.2}
\log c_\cuh = -\sum_i \frac{h_i}{2} \Big( \log \frac{h_i}{2}-1\Big)
- \oh \log \Del (h), ~~~~\Del(h) = \prod_{i<j} (h_i-h_j). 
\eeq
Finally, if $\chi_\cuh$ denotes the character associated with $\cuh$,
\beq{3.3}
R_\cuh (\a) = \int \d A \; \chi_\cuh(A) \; \exp M \Big( -\oh \tr A^2 + 
\frac{\a}{4} \tr A^4\Big).
\eeq
It is now possible to perform a double saddle point expansion of \rf{3.1}
and \rf{3.3}. In order to describe the formalism let us introduce the 
notation 
\beq{3.a}
\Re f(z) \equiv \frac{f(z+i0)+f(z-i0)}{2},~~~~~
\Im f(z) = \frac{f(z+i0)-f(z-i0)}{2}.
\eeq
This notation is useful when $f(z)$ has cuts. The saddle point expansion 
assumes the existence of an eigenvalue density $\tilde{\rho}(\l)$, 
or equivalently a resolvent associated with the matrix integral \rf{3.3}:
\beq{3.4}
\om (\l) = \frac{1}{M} \sum_k \frac{1}{\l -\l_k},~~~~
-\pi i \tilde{\rho}(\l)= \Im \om (\l),
\eeq
and (after rescaling $h \to h/M$)
a density of highest weights $\rho(h)$, or the corresponding
``resolvent'' $H(h)$:
\beq{3.5}
H(h) =\frac{1}{M} \sum_{k} \frac{1}{h-h_k},~~~~~-\pi i \rho(h) = \Im H(h).
\eeq

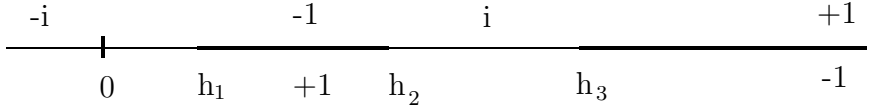
\begin{figure}[t]
\begin{center}
\setlength{\unitlength}{3947sp}%
\begingroup\makeatletter\ifx\SetFigFont\undefined%
\gdef\SetFigFont#1#2#3#4#5{%
  \reset@font\fontsize{#1}{#2pt}%
  \fontfamily{#3}\fontseries{#4}\fontshape{#5}%
  \selectfont}%
\fi\endgroup%
\begin{picture}(5498,653)(889,-2521)
\thinlines
{\put(901,-2161){\line( 1, 0){5400}}
}%
\thicklines
{\put(2101,-2161){\line( 1, 0){1200}}
}%
{\put(4501,-2161){\line( 1, 0){1800}}
}%
{\put(1508,-2093){\line( 0,-1){128}}
}%
\put(2701,-2011){\makebox(0,0)[lb]{\smash{\SetFigFont{12}{14.4}{\rmdefault}{\mddefault}{\updefault}{-1}%
}}}
\put(2701,-2461){\makebox(0,0)[lb]{\smash{\SetFigFont{12}{14.4}{\rmdefault}{\mddefault}{\updefault}{+1}%
}}}
\put(2101,-2461){\makebox(0,0)[lb]{\smash{\SetFigFont{12}{14.4}{\rmdefault}{\mddefault}{\updefault}{h}%
}}}
\put(3301,-2461){\makebox(0,0)[lb]{\smash{\SetFigFont{12}{14.4}{\rmdefault}{\mddefault}{\updefault}{h}%
}}}
\put(2206,-2483){\makebox(0,0)[lb]{\smash{\SetFigFont{8}{9.6}{\rmdefault}{\mddefault}{\updefault}{1}%
}}}
\put(3428,-2521){\makebox(0,0)[lb]{\smash{\SetFigFont{8}{9.6}{\rmdefault}{\mddefault}{\updefault}{2}%
}}}
\put(1051,-2026){\makebox(0,0)[lb]{\smash{\SetFigFont{12}{14.4}{\rmdefault}{\mddefault}{\updefault}{-i}%
}}}
\put(3893,-2033){\makebox(0,0)[lb]{\smash{\SetFigFont{12}{14.4}{\rmdefault}{\mddefault}{\updefault}{i}%
}}}
\put(5993,-2003){\makebox(0,0)[lb]{\smash{\SetFigFont{12}{14.4}{\rmdefault}{\mddefault}{\updefault}{+1}%
}}}
\put(6023,-2408){\makebox(0,0)[lb]{\smash{\SetFigFont{12}{14.4}{\rmdefault}{\mddefault}{\updefault}{-1}%
}}}
\put(4478,-2453){\makebox(0,0)[lb]{\smash{\SetFigFont{12}{14.4}{\rmdefault}{\mddefault}{\updefault}{h}%
}}}
\put(4606,-2491){\makebox(0,0)[lb]{\smash{\SetFigFont{8}{9.6}{\rmdefault}{\mddefault}{\updefault}{3}%
}}}
\put(1486,-2468){\makebox(0,0)[lb]{\smash{\SetFigFont{12}{14.4}{\rmdefault}{\mddefault}{\updefault}{0}%
}}}
\end{picture}
\end{center}
\caption[cut]{The cut structure of $r(h)$ in the complex $h$-plane.}
\label{cut}
\end{figure}

In \cite{kz} the double saddle point expansion is analyzed in the case 
$\a_1 \equ \a_2\equ \a$. 
The density $\rho(h)$ was assumed to 
be different from zero in the interval $[0,h_2[$, and equal 
to 1 in the interval $[0,h_1]$, where $0<h_1<h_2$. 
Further, for a given eigenvalue 
distribution $\l_k$ of the matrix $A$ coming from the 
saddle point of \rf{3.3} one can define a function $L(h)$,
with same cut as $H$ by
\beq{3.6}
\Re L(h_j) = \frac{2}{M}  \frac{\prt}{\prt h_j} \log \chi_\cuh (A(\l_k)).
\eeq 
The analysis of \cite{kz}  
shows that $L(h)= H(h)+F(h)$ where 
$F(h)$ is analytic on the cut of $H(h)$ but has an additional 
cut  $[h_3,\infty[$ where 
\beq{3.7}
2 \Re L(h) = \log \frac{h}{\a} + H(h).
\eeq
It can now be shown that the function 
$D(h)\equ 2L(h)-H(h) -3\log h + \log (h-h_1)$ only has square root type 
cuts on $[h_1,h_2]$ and $[h_3,\infty[$ and on these cuts satisfies the 
following equations:
\bea
\Re D(h) &=& \log \frac{h-h_1}{\b h^2},~~~h\in I_0= [h_1,h_2]  \label{3.8}\\
\Re D(h) &=& \log \frac{h-h_1}{\a h^2},~~~h \in I_1 = [h_3,\infty[. \label{3.9}
\eea
Eqs.\ \rf{3.8}--\rf{3.9} constitute a standard Hilbert problem
and the inversion formula is unique \cite{book}.  
The function holomorphic 
in the plane with cuts $I_0$ and $I_1$ is  given by 
\bea
D(h) &=& \log \frac{h-h_1}{\b h^2} -  \frac{\log \b/\a}{i\pi}\, r(h) 
\int_{h_3}^\infty {\d h'} \, \frac{1}{(h-h') r(h')} \label{g7} \\
&&+ r(h) \int_{-\infty}^{h_1} {\d h'} \frac{1}{(h-h') r(h')}-
2  r(h) \int_{-\infty}^{0} {\d h'} \frac{1}{(h-h') r(h')}
\no
\eea
where 
\beq{g2}
r(h) = \sqrt{(h-h_1)(h-h_2)(h-h_3)}
\eeq
and where we have chosen the cut structure shown in Fig.\ \ref{cut}.
Following \cite{book} the meaning of $r(h')$ on the cut is $r(h'\pl i0)$,
i.e.\ the function on the ``left side'' of the cut. The integrals can 
be expressed in terms of standard elliptic functions. However, 
we do not need the explicit expressions here.

From $D(h)$ we can derive 
the expression for $\rho(h)$ which is 
\beq{g8}
\rho(h) = -\frac{\Im H(h)}{i\pi} =  -\frac{\Im D(h)}{i\pi},~~~~h \in ]h_1,h_2[.
\eeq
We have ($h$ always $h\pl i0$ if ambiguities) :
\bea
\rho(h) &=&  \frac{-r(h)}{\pi} \int_{-\infty}^{h_1} 
\frac{\d h'}{(h-h') ir(h')}-
\frac{(-2 r(h))}{\pi} \int_{-\infty}^{0} \frac{\d h'}{(h-h')i r(h')}+
\no\\
&&
+\frac{\log \b/\a}{\pi^2}\,(-r(h))\int_{h_3}^\infty \frac{\d h'}{(h-h') r(h')}.
\label{g9}
\eea

Note that the derivative of $D(h)$ and $\rho(h)$ 
with respect to $h_i$ are elementary 
functions of $h$. For instance, differentiating with respect
to $h_3$ we have
\beq{expand}
\frac{\prt D(h)}{\prt h_3} = \frac{i r(h)}{h-h_3}\; \frac{\prt W}{\prt h_3}
= \frac{i r(h)F_3}{2(h-h_3)}  = -i\pi \frac{\prt \rho(h)}{\prt h_3},
\eeq
where the last equality is valid for $h \in I_0=[h_1,h_2]$ and 
where $W(h_1,h_2,h_3)$ and $F_3(h_1,h_2,h_3)$ are  
defined below (eqs.\ \rf{bound1} and \rf{e5}). Thus we can write
\bea\label{expanD}
D(h;h_3+\del h_3) &=& D(h;h_3)+ \del h_3 \, 
\frac{i r(h)F_3(h_3)}{2{(h-h_3)}}\\
&&+ {(\del h_3)^2}
\frac{i r(h)}{4(h-h_3)} 
\left(F'_3(h_3)+\frac{F_3(h_3)}{2(h-h_3)}\right)+\cdots
\no
\eea
and similarly for $\rho(h)$. The function $F_3(h_1,h_2,h_3)$ is a sum 
of elliptic integrals.

\subsection{Boundary conditions}\label{boundary}

The starting formula for $D(h)$ is \rf{g7}.
The general large-$h$ behaviour of this function is 
\beq{II.4}
c_1 h^{1/2} - \log (-\a h) + c_2 h^{-1/2} + O(1/h).
\eeq
However, according to the analysis in \cite{kz} 
$c_1 \equ 0$ and $c_2 \equ -(-\a)^{-1/2}$. This gives 
two boundary conditions for the constants $h_1,h_2,h_3$ which 
appear in $r(h)$ and thus in $D(h)$. 
The coefficients $c_1$ and $c_2$  
can be identified by expanding the integrand in 
\rf{g7} in powers of $1/h$ and one obtains the boundary 
conditions:
\beq{II.5}
c_1 =i W(h_1,h_2,h_3) =0,~~~~
c_2 = i \Omega = \frac{i}{\sqrt{\a}},
\eeq
where we have used the first of the equations \rf{II.5} to 
simplify the second, and where  
\beq{bound1}
W(h_1,h_2,h_3) = \frac{\log \b/\a}{\pi} 
 \int_{h_3}^\infty \frac{\d h'}{r(h')} 
+\int_{-\infty}^{h_1} \frac{\d h'}{ir(h')}-
2 \int_{-\infty}^{0} \frac{\d h'}{ir(h')},
\eeq
\beq{e4}
\Omega(h_1,h_2,h_3)= 
 -\frac{\log \b/\a}{\pi} 
 \int_{h_1}^{h_2} \frac{h'\d h'}{r(h')} 
+\int_{h_2}^{h_3} \frac{h'\d h'}{ir(h')}+
2 \int_{0}^{h_1} \frac{h'\d h'}{ir(h')}.
\eeq
We will not need the explicit expressions for these integrals.
The final boundary condition is 
\beq{bound3}
\int_{h_1}^{h_2} \d h \;\rho(h) = 1-h_1,
\eeq
where $\rho(h)$ is given by \rf{g9}.

For given $(\a,\b)$, we can in principle solve 
the three boundary conditions \rf{II.5} and \rf{bound3}
for $(h_1,h_2,h_3)$, thereby obtaining a solution of the 
matrix integral. Equivalently, the three conditions
define locally a map from the $\beta$-$\alpha$-plane to
a two-dimensional hypersurface in the parameter space of
the $h_i$.  Critical regions of the free energy $F$ of the
model are associated with singularities of the inverse
map of an independent subset of the $h_i$'s, say, 
$(h_2,h_3)\mapsto (\a (h_2,h_3),\b (h_2,h_3))$, which
lead to singularities of $F (\a,\b)$ upon substitution.

\subsection{The critical line}

The generic behaviour of $D(h)$ when $h \to h_3$ is clearly
$(h-h_3)^{1/2}$, simply coming from the term $r(h)$ in the 
representation \rf{g7}. However, this behaviour can change
to $(h-h_3)^{3/2}$
along a curve $\a_c(\b)$ in the $(\b,\a)$ coupling constant plane.
According to \cite{kz} this is the critical line of phase 
A of the ABAB matrix model and according to \cite{ajlv} this is 
where the continuum limit of 3d gravity should be found.
Similarly the criticality in the B phase is derived from the behaviour
of $D(h)$ when $h \to h_2$, where a generic behaviour $(h-h_2)^{1/2}$ changes
to $(h-h_2)^{3/2}$.

We now study the change of $(h_1,h_2,h_3)$ as $\a$ and $\b$ change
infinitesimally. For simplicity we first present the 
result when $\a/\b$ is constant.

Let us first identify the coefficient of $\sqrt{h-h_3}$ in $D(h)$.
Using \rf{expand} in the expression for  $D(h)$ the coefficient
can be written as 
\beq{e5}
ir_0(h_3) F_3(h_1,h_2,h_3), ~~~~r_0(h) \equiv \sqrt{(h-h_1)(h-h_2)},
\eeq
One has 
\beq{e7}
\frac{\prt W(h_1,h_2,h_3)}{\prt h_3} = \oh F_3(h_1,h_2,h_3).
\eeq
We define $F_1$ and $F_2$ similarly to $F_3$ and have relations
like \rf{e7}.
From the boundary conditions \rf{II.5} it follows that the variation 
of $h_1,h_2,h_3$ as $\a,\b$ change with the ratio $\a/\b$ fixed satisfy
\beq{nbound1}
F_1 \del h_1+
F_2 \del h_2+ F_3 \del h_3= 0.
\eeq
\beq{nbound2}
h_1 F_1 \del h_1 \pl h_2 F_2\del h_2 \pl 
h_3F_3\del h_3 = 
-2\frac{\del \a}{\a^{3/2}}.
\eeq

The final boundary condition involves the density. Since $\rho(h_1)\equ 1$
and $\rho(h_2) \equ 0$ the variation of \rf{bound3} just becomes
\beq{e10}
\int_{h_1}^{h_2} \d h \; \Big(
\frac{\prt \rho}{\prt h_1} \del h_1+
\frac{\prt \rho}{\prt h_2} \del h_2+
\frac{\prt \rho}{\prt h_3} \del h_3\Big)= 0,
\eeq
which can be written, using (\ref{expand}), as 
\beq{nbound3}
E_1F_1 \del h_1+E_2F_2 \del h_2+E_3F_3 \del h_3 = 0,
\eeq
where 
\beq{e13} 
\int_{h_1}^{h_2} \frac{\d h \; r(h)}{h_i-h} = E_i,~~~~i= 1,2,3
\eeq
are elliptic integrals.
It is straightforward to repeat the derivation in the case where  
the ratio $\a/\b$ is not assumed constant, 
leading to the set of equations
\bea
&&R_1 \biggl(\frac{\del\a}{\a}-\frac{\del\b}{\b}\biggr)+
F_1 \del h_1+F_2 \del h_2+ F_3 \del h_3= 0,\nonumber\\
&&R_2 \biggl(\frac{\del\a}{\a}-\frac{\del\b}{\b}\biggr)+
\frac{\del\a}{\a^{3/2}}+
h_1 F_1 \del h_1 \pl h_2 F_2\del h_2 \pl h_3F_3\del h_3 = 0,
\nonumber\\
&&R_3 \biggl(\frac{\del\a}{\a}-\frac{\del\b}{\b}\biggr)+
E_1 F_1 \del h_1 \pl E_2 F_2\del h_2 \pl E_3F_3\del h_3 = 0,
\label{total}
\eea
for the linear variations, 
where the functions $R_i(h_1,h_2,h_3)$ are given by
\bea
&&R_1=-\frac{2}{\pi}\int_{h_3}^{\infty}\frac{dh'}{r(h')},
\;\;\;
R_2=\frac{2}{\pi}\int_{h_1}^{h_2}\frac{h'dh'}{r(h')},
\nonumber\\
&&R_3=\frac{2}{\pi}\int_{h_1}^{h_2}dh\ r(h)
\int_{h_3}^{\infty}\frac{dh'}{(h-h')r(h')}.\label{rdef}
\eea

Without loss of generality, we can now choose $h_2$ and $h_3$ as
independent variables on the 2d hypersurface. Using \rf{total}
to eliminate $\del h_1$, one can derive the associated linear map
between the remaining variables, 
\beq{2dmat}
\begin{pmatrix}
\del\a /\a\cr \del\b /\b
\end{pmatrix} =
\begin{pmatrix}
X_1 F_2& Y_1 F_3\cr
X_2 F_2& Y_2 F_3
\end{pmatrix}
\begin{pmatrix}
\del h_2\cr \del h_3
\end{pmatrix},
\eeq
where $X_i$ and $Y_i$ are easily calculable functions of the $R$'s,
$E$'s and $h$'s.
This transformation is in general well defined, unless either
$F_3\equ 0$ or $F_2\equ 0$, making the Jacobian vanish.
We already know that these two equations define a critical
line in coupling-constant space, the former giving rise to
phase A, and the latter to phase B. They meet in a single
critical point with $F_3\equ F_2\equ 0$.

In determining a continuum limit of the matrix model
related to three-dimensional quantum gravity, we are
interested in the relation between $(\b,\a)$ and $(h_2,h_3)$
along specific curves as they approach a point on 
the critical line.
Note first that along generic curves and away from the
critical line, by virtue of \rf{2dmat} all variations will
be of the same order, namely
\beq{solu1}
\del \a\sim\del \b\sim \del h_2\sim \del h_3.
\eeq
This behaviour changes when the critical line is approached
from an infinitesimal distance. As can be seen from
\rf{2dmat}, any such curve whose tangent does not coincide
with that of the critical line in phase A at 
their mutual intersection point
has vanishing derivatives $\partial \a/\partial h_3$ and 
$\partial \b/\partial
h_3$ there, by continuity. This means that although 
the variations $\del\a$ and $\del\b$ are of the same order
as $\del h_2$ (and therefore also $\del h_1$), their relation with 
$\del h_3$ is of higher order, indeed,
\beq{solu2}
\del \a\sim \del\b\sim  \del h_2 \sim (\del h_3)^2.
\eeq
A completely analogous statement holds in phase B of the 
model with $h_2$ and $h_3$ interchanged, since the
critical line is defined by $F_2\equ 0$ there, instead of
$F_3\equ 0$. The qualitative difference between the
two phases will only become apparent in the discussion
of the asymmetric case below.

An alternative way of approaching a point on the critical
line that turns out to be relevant for quantum gravity is
by moving {\it along} the line itself. This case is analyzed
most transparently by adding (in phase A) the constraint
$F_3\equ 0$ and expanding it along with \rf{II.5} and \rf{bound3}.
One verifies by a short calculation that the first-order
variations in this case behave according to \rf{solu1}.

\section{The asymmetric case \boldmath{$\a_1 \neq \a_2$}.}\label{asym}

As mentioned earlier, the construction of the 
transfer matrix  requires that we perturb away from $\a_1\equ \a_2$.
Let us  discuss the general structure of the matrix model with 
$\a_1 \neq \a_2$ (as explained above  we will only 
need an infinitesimal perturbation away from $\a_1\equ \a_2$
in the continuum limit)\footnote{
As already mentioned an explicit solution of the asymmetric ABAB model 
has been published while this manuscript was being completed \cite{newPZJ}.}. 
The main difference in the analysis of the 
matrix model with $\a_1 \neq \a_2$ compared to the situation 
$\a_1\equ \a_2 \equ \a$ is that the saddle point solution 
involves two eigenvalue densities $\trho_{1}(\l) $ and 
$\trho_{2}(\l)$ corresponding to the two one-matrix integrals
\rf{3.3} with $\a\equ \a_1$ and $\a\equ \a_2$. Similarly, we 
will have two functions  $L_1(h)$ and $L_2(h)$ corresponding to 
\rf{3.6} since the eigenvalue densities $\trho_1(\l)$ and $\trho_2(\l)$ 
appear via the matrix $A(\l)$ in \rf{3.6}. 
On other hand we have only one density $\rho(h)$ 
coming from the saddle point of \rf{3.1}. In order to solve the 
saddle point equations it is natural to follow the same strategy as 
in \cite{kz} and make an educated guess about the analytic structure 
of the functions involved and then show the self-consistency of the 
solution. Since we have two functions $L_i(h)$, associated with the 
same $\rho(h)$ but different $\trho(\l)$'s, and the cut of $L(h)$ 
from $[h_3,\infty[$ can be traced to the saddle point equation 
for $\trho(\l)$ (see \cite{kz} for a discussion), 
it is natural to assume that $L_1(h)$ and $L_2(h)$ 
have a cut from $[0,h_2]$ (with $\rho(h) \equ 1$ in $[0,h_1]$),
and that they have separate cuts $[h_3^{(1)},\infty[$ and  
$[h_3^{(2)},\infty[$. In the case where $\b \equ 0$ this structure 
is indeed realized.

We can now write down the generalization of \rf{3.8}-\rf{3.9}
\bea
\Re [{D}_1(h) ] & = & \log \frac{h-h_1}{h^2 \beta}+\Re [f(h)] 
~~~~~   \forall h \in I_0 \equiv [h_1,h_2] \label{S1a}\\
\Re [ {D}_1(h)] & = & \log \frac{h-h_1}{h^2 \alpha_1}  
~~~~~~~~~~~~~~~~~~ \forall h \in I_1\equiv [h_3^{(1)},\infty[  \label{S1b}
\eea
and 
\bea
\Re[{D}_2(h)] & = & \log \frac{h-h_1}{h^2 \beta}-\Re [f(h)] 
~~~~~ \forall h \in  I_0 \equiv [h_1,h_2]\label{S2a}\\ 
\Re[{D}_2(h)] & = & \log \frac{h-h_1}{h^2 \alpha_2} 
~~~~~~~~~~~~~~~~~~ \forall h \in I_2 \equiv [h_3^{(2)},\infty[. \label{S2b} 
\eea
In \rf{S1a}-\rf{S2b} the $D$'s are related to the $L$'s and the 
function $H$ as below eq.\ \rf{3.7}, that is,
\beq{3.20}
D_i(h) = 2L_i(h) -H(h) -3\log h +\log(h-h_1),~~~~i=1,2,
\eeq
where the subtractions of the log's are made to ensure that the functions 
$D_i$'s have square root cuts. As in the case of a single $\a$, we assume 
that $L_i(h)\equ F_i(h) + H(h)$ where $F_i(h)$ is analytic on the 
cut $I_0$ of $H(h)$.\footnote{The $F_i$ should not be confused
with the functions of the same name defined in \rf{e7} above.}
The function
\beq{3.21}
f(z) \equiv F_1(z)-F_2(z) = L_1(z)-L_2(z)
\eeq
is at this point unknown, but we can write $\Re [f(z)] \equ f(z)$ on
$I_0$. If we assume $f(z)$ is known on $I_0$, eqs.\ \rf{S1a}-\rf{S1b} 
and \rf{S2a}-\rf{S2b} are standard singular integral equations 
of the Hilbert type and can readily be solved and one can write
\bea
{D}_{1}(z) & = & D^{kz}_{1}(z) + r_{1}(z) 
\oint_{I_{0}} \frac{dt}{2 \pi i} 
\frac{f(t) }{(z-t) r_{1}(t)} \label{DD1a}\\
{D}_{2}(z) & = & D^{kz}_{2}(z) - r_{2}(z) 
\oint_{I_{0}} \frac{dt}{2 \pi i} 
\frac{f(t) }{(z-t) r_{2}(t)}\label{DD1b}
\eea
where $D^{kz}_{1,2}(z)$ are given by formula \rf{g7} with 
$(\a,h_3)\equ (\a_1,h_3^{(1)})$ and $(\a,h_3) \equ (\a_2,h_3^{(2)})$, 
respectively, and
\beq{ridef}
r_i(h)=\sqrt{(h-h_1)(h-h_2)(h-h_3^{(i)})}.
\eeq
Furthermore, we have
\beq{im}
\Im [D_k(h)] =-i \pi \rho(h),~~~~k=1,2.
\eeq
Therefore, the 
``imaginary'' parts of eqs. (\ref{DD1a}) and \rf{DD1b} are
\bea
i \pi \rho(h) & = &  i \pi \rho^{kz}_{1}(h) -
r_{1}(h) 
\Pint_{I_{0}} \frac{dt}{ \pi i} 
\frac{f(t)}{(t-h) r_{1}(t)} \, \qquad \forall h \in I_0  
\label{density1}\\
i \pi \rho(h) & = &  i \pi \rho^{kz}_{2}(h) +
r_{2}(h) 
\Pint_{I_{0}} \frac{dt}{ \pi i} 
\frac{f(t)}{(t-h) r_{2}(t)} \, \qquad \forall h \in I_0 
\label{density2}
\eea
where $\Pint$ is the principal value of the integral. Eqs.\ 
(\ref{density1})--\rf{density2} determine $f(z)$ and $\rho(h)$ in terms of 
the densities $\rho^{kz}_1,\rho^{kz}_2$ (eq.\ \rf{feq1}),
corresponding to $\a\equ  \a_1$, $h_3 \equ h_3^{(1)}$ and 
$\a \equ \a_2$ , $h_3\equ h_3^{(2)}$. 
We can obtain an equation for $f(z)$ by subtracting
\rf{density1} and \rf{density2}, leading to
\beq{feq}
i \pi (\rho^{kz}_1(h)-\rho^{kz}_2(h))=
\Pint_{I_0} \, \frac{dt \; f(t)}{ \pi i (t-h)} 
\left( \frac{r_1(h)}{r_1(t)}+ \frac{r_2(h)}{r_2(t)}
\right)  .
\eeq 

\subsection{Uniqueness of the solution}

Let us discuss the solution of \rf{feq} which is a singular integral 
equation. In order to bring it into a standard form of singular 
integral equations, and for convenience of later applications, 
we introduce the notation
\beq{feq0}
\Del \rho^{kz}(h) \equiv \frac{i \pi}{2}\, 
\Big(\rho^{kz}_1(h)-\rho^{kz}_2(h)\Big),
\eeq
and 
\beq{feq0a}
(t-h)\,k(h,t)\equiv \oh 
\sqrt{\frac{t-h_3}{h-h_3}}
\left(\sqrt{\frac{h-h_3^{(1)}}{t-h_3^{(1)}}}+ 
\sqrt{\frac{h-h_3^{(2)}}{t-h_3^{(2)}}}-2\sqrt{\frac{h-h_3}{t-h_3}}\right),
\eeq
where from now on $h_3$ will always refer to the average  
\beq{fex}
h_3= \oh(h_3^{(1)}+h_3^{(2)}),
\eeq
and the function $r(h)$ will refer to \rf{g2} with $h_3$ given 
by \rf{fex}.

The function $k(h,t)$ is regular at $h\equ t$. Let us further introduce 
\beq{feq0b}
\Drhro(h) \equiv  \frac{\Del \rho^{kz}(h) }{r(h)},
~~~~~~~~\fro(t) \equiv \frac{f(t)}{r(t)}.
\eeq
We can now write eq.\ \rf{feq} as  
\beq{feq1}
\PintI \dt\; \frac{\fro(t)}{t-h} + 
\int_{I_0} \dt \; k(h,t) \fro(t) = \Drhro (h),
\eeq
where only the first integral is singular.
The so-called {\it dominant part} of this singular integral equation
is given by 
\beq{feq2}
\Pint_{I_0} \, \dt  \frac{\tilde{\fro}(t)}{t-h}= \Drhro (h).
\eeq
This equation has precisely one  zero mode, namely,
\beq{feq3}
\tilde{\fro}(t) = 
\frac{1}{r_0(t)}\PintI \dh \; \frac{r_0(h) \Drhro(h) }{h-t} + 
\frac{C}{r_0(t)},
\eeq
where $r_0(t) = \sqrt{(t-h_1)(t-h_2)}$.
Expressed in terms of $\tilde f(t)$ we have
\beq{feq4}
\tilde{f}(t) = \sqrt{t-h_3} \left(\PintI \dh \; 
\frac{ \Del \rho^{kz}(h)}{(h-t)\sqrt{h-h_3}} 
+C\right).
\eeq
By moving the $k$-term in Eq.\ \rf{feq1} to the right-hand side, 
we can repeat the 
steps leading to \rf{feq4}, and moving the $k$-term back to the 
left-hand side we finally obtain
\beq{feq5}
f(t) + \int_{I_0} \d s N(t,s)\, f(s) = \sqrt{t-h_3}
\left(\PintI \dh \frac{\Del \rho^{kz}(h) }{(h-t)\sqrt{h-h_3}}+
C\right),
\eeq
where the kernel $N(t,s)$ is a {\it Fredholm kernel},
\beq{feq6}
N(t,s) = -\frac{\sqrt{t-h_3}}{r(s)}
\PintI \frac{\d h}{\pi^2} \;
\frac{k(h,s) r_0(h)}{h-t}.
\eeq
In general the solution to the Fredholm equation \rf{feq5} will be 
unique \cite{book}. 
{\it We thus have a one-parameter family of solutions $f_C(t)$}.

In order to determine the {\it four} parameters 
$h_1,h_2,h^{(1)}_3,h^{(2)}_3$ and the constant $C$
we need {\it four} boundary conditions and one more 
condition, which in this case will be the normalization condition 
for $\rho$ (which appeared above in formulas \rf{density1}
and \rf{density2}),
\beq{fifth}
h_1+\int_{h_1}^{h_2} \rho(t) dt -1=0.
\eeq
The boundary conditions are again obtained by the requirement 
that the large-$h$ asymptotics of $D_j(h)$ contain no $h^{1/2}$ term 
while the coefficient of the $h^{-1/2}$ term equal $(-\a_j)^{-1/2}$.
This results in four boundary conditions
\beq{cond1}
{\cal W}_j := 
W_j -(-1)^j \int_{I_0} 
\frac{dt}{\pi} \frac{f(t)}{r_j(t)} = 0,
\eeq
\beq{cond2}
{\cal O}_j:= \Omega_j-(-1)^j
\int_{I_0} \frac{dt}{\pi} \frac{ f(t) t}{r_j(t)}-
\frac{1}{\sqrt{\alpha_j}}=0,
\, 
\eeq
where $j\equ 1,2$, $W_j\equ W(h_1,h_2,h_3^{(j)})$
and $\Omega_j \equ \Omega (h_1,h_2,h_3^{(j)})$.

The set of constraints \rf{fifth} \rf{cond1}, \rf{cond2} implies 
that in the asymmetric case the theory
is defined on a three-dimensional hypersurface of the eight-dimensional
parameter space $(\a_1,\a_2,\b,C,h_1,h_2,h_3^{(1)},h_3^{(2)})$. 
Singular points
of the map between the parameters $(\a_1,\a_2,\b)$ and some
other parametrization of this hypersurface (say, in terms of 
$(h_2,h_3^{(1)},h_3^{(2)})$, after eliminating $C$ and $h_1$) 
will correspond to critical points of the theory.  

In analogy with the symmetric case we expect criticality to
be related to a change in the behaviour of the $D_j(h)$ when
$h$ approaches one of the $h_3^{(j)}$ (phase A) or $h_2$ (phase B). 
In the first case, this is signalled by the vanishing of one
of the coefficients ${\cal F}_3^{(j)}$ in the relevant expansion
\beq{Cgen}
D_j(h)={\cal F}_3^{(j)}\sqrt{h-h_3^{(j)}}+{\rm higher\ order},
\eeq
with
\beq{f3coeff}
{\cal F}_3^{(j)}:=F_3^{(j)}-\int_{I_0}\frac{dt}{\pi} \frac{f(t)}
{(h_3^{(j)}-t)r_j(t)}
\eeq
where $F_3^{(j)}\equ F_3(h_3\equ h_3^{(j)})$.
The two critical hypersurfaces in phase A defined by 
${\cal F}_3^{(j)}$, $j=1,2$, intersect along a critical line which
coincides with the critical line of the symmetric model.

In phase B, there is a single smooth critical hypersurface
defined by the simultaneous vanishing of the two functions
${\cal F}_2^{(j)}$ (defined similar to \rf{f3coeff})
appearing in the expansions
\beq{Cgen1}
D_j(h)={\cal F}_2^{(j)}\sqrt{h-h_2}+{\rm higher\ order}.
\eeq
The simultaneous change in behaviour of ${\cal F}_2^{(1)}$
and ${\cal F}_2^{(2)}$ can be traced to the fact that
the non-analyticity at $h_2$ is due entirely to $H(h)$ (or
$\rho (h)$) which are common to $D_1(h)$ and $D_2(h)$.
The critical hypersurface defined by ${\cal F}_2^{(j)}\equ 0$
intersects both hypersurfaces ${\cal F}_3^{(1)}\equ 0$ and
${\cal F}_3^{(2)}\equ 0$ in critical lines, and there is a
single (tricritical) point where all of the three hypersurfaces
meet.

One can now proceed as in the symmetric case by expanding the
five constraint equations to first (or any desired) order,
and eliminating the dependent variations $\del h_1$ and
$\del C$, say. The three remaining first-order equations can be
solved for $\del\a_1$, $\del\a_2$ and $\del\b$ and put in
matrix form,
\beq{3dmat}
\begin{pmatrix}
\del\a_1 /\a_1\cr 
\del\a_2/\a_2\cr
\del\b /\b
\end{pmatrix} =
\begin{pmatrix}
{\cal X}_{11} {\cal F}_3^{(1)}+{\cal Z}_{11} &
{\cal X}_{12} {\cal F}_3^{(2)}+{\cal Z}_{12} &
{\cal X}_{13} {\cal F}_2^{(1)}+{\cal Z}_{13} \cr
{\cal X}_{21} {\cal F}_3^{(1)}+{\cal Z}_{21} &
{\cal X}_{22} {\cal F}_3^{(2)}+{\cal Z}_{22} &
{\cal X}_{23} {\cal F}_2^{(1)}+{\cal Z}_{23} \cr
{\cal X}_{31} {\cal F}_3^{(1)}+{\cal Z}_{31} &
{\cal X}_{32} {\cal F}_3^{(2)}+{\cal Z}_{32} &
{\cal X}_{33} {\cal F}_2^{(1)}+{\cal Z}_{33} 
\end{pmatrix}
\begin{pmatrix}
\del h_3^{(1)}\cr
\del h_3^{(2)}\cr \del h_2
\end{pmatrix},
\eeq
where the functions ${\cal X}_{ij}$ and ${\cal Z}_{ij}$ 
are regular and generically non-vanishing.
Moreover, it can be shown that the three functions 
${\cal Z}_{i1}$ vanish at points
$h\equ h_3^{(1)}$ of the critical surface defined
by ${\cal F}_3^{(1)}\equ 0$, by virtue of the fact that
the derivative $\partial f/\partial h_3^{(1)}$ at such
points is proportional to the zero-mode of $f$.
(Analogous statements hold for ${\cal Z}_{i2}$ and 
${\cal Z}_{i3}$.) 
The situation is therefore very similar to what happened
in the symmetric case, cf. \rf{2dmat}. Namely, the rank
of the Jacobian of the coordinate transformation
$(\a_1,\a_2,\b)\mapsto (h_3^{(1)},h_3^{(2)}, h_2)$ drops
by one when we go to one of the critical surfaces.

Expressed in terms of the linear variations, this means
that away from any of the critical surfaces one has
\beq{solu3}
\del \a_1\sim\del \a_2\sim\del \b\sim \del h_2\sim 
\del h_3^{(1)}\sim\del h_3^{(2)},
\eeq
and that this behaviour changes when a critical surface,
say, ${\cal F}_3^{(1)}\equ 0$ is approached, to
\beq{solu4}
\del \a_1\sim\del \a_2\sim\del\b\sim \del h_2
\sim (\del h_3^{(1)})^2   \sim\del h_3^{(2)}.
\eeq
Similarly, it follows that if the critical (symmetric) 
line of phase A is approached along any path that is
not tangential to either of the critical surfaces
${\cal F}_3^{(j)}$, $j\equ 1,2$, the variations will
behave like
\beq{solu5}
\del \a_1\sim\del \a_2\sim\del\b\sim \del h_2 \sim(\del 
h_3^{(1)})^2 \sim (\del h_3^{(2)})^2.
\eeq

As before in the symmetric case, if one chooses a linear
variation tangential to one of the critical hypersurfaces,
say, ${\cal F}_3^{(1)}\equ 0$, this imposes a linear
relation between $\del h_3^{(1)}$, $\del h_3^{(2)}$ and
$\del h_2$, and at the same time forces $\del h_3^{(1)}$ 
to be of the same order as the other two, leading again
to the behaviour \rf{solu3}. The same is true if we
consider a variation tangent to the critical line in phase
A. This case is analyzed easily by adjoining the two
additional conditions ${\cal F}_3^{(j)}\equ 0$ to the
other boundary conditions, and expanding them to first
order.

\section{Expanding around \boldmath{$\a_1= \a_2$}}\label{expander}

In applying the matrix model to 3d quantum gravity, we do not 
need the complete explicit solution of the asymmetric ABAB model,
but only infinitesimal variations 
around symmetric points with $\a_1\equ\a_2$ and
$h_3^{(1)}\equ h_3^{(2)}$. In this context, it is convenient
to work with the symmetrized and anti-symmetrized variables
\beq{b-a}
\a=\sqrt{\a_1\a_2},\;\;\; \tilde\a=\sqrt{\frac{\a_1}{\a_2}},\;\;\;
h_3= \frac{h_3^{(1)}+h_3^{(2)}}{2},\;\;\;
\tilde h_3= \frac{h_3^{(1)}-h_3^{(2)}}{2}.
\eeq
Infinitesimal linear variations around any given
point $(\a^{(0)},
\tilde\a^{(0)},\b^{(0)})$ or the corresponding 
$(h_3^{(0)},\tilde h_3^{(0)},h_2^{(0)})$ will be denoted by
$\delta\a$, $\delta\tilde\a$, $\delta\b$, $\delta h_3$,
$\delta\tilde h_3$ and $\delta h_2$. In what follows, we
will study further the relations between these variations which
follow from the conditions \rf{fifth}, \rf{cond1} and \rf{cond2}, 
since they determine the continuum physics of the
model.

\subsection{Finding {$f(t)$} and {$D_{1,2}(h)$}}

The discussion in this section will concentrate on phase
A, which is the most relevant one for the quantum-gravitational
application. We will comment briefly at the end on what happens 
in phase B.

In order to make maximal use of the relations we already
derived for the symmetric case, it is convenient
to perform a partial expansion of the relevant functions
around points in the symmetric plane (characterized
by $\tilde \a \equ 1$, or, equivalently, $\tilde h_3\equ 0$).
We will Taylor-expand around $\tilde h_3^{(0)}\equ 0$ and 
simultaneously allow for a finite shift to 
$\tilde \a\equ {\rm e}^\varepsilon$ away from the symmetric value
$\tilde \a^{(0)} \equ 1$.

Recall now the integral equation \rf{feq5} which determines $f(t)$.
Under the variation introduced in the last paragraph, because of
symmetry in $h_3^{(1)}$ and $h_3^{(2)}$, the function 
$k(t,h)$ has an expansion in even powers of $\del\tilde h_3$,
\beq{kth}
k(t,h) = (\del\tilde h_3)^2 k_1(t,h) + (\del\tilde h_3)^4k_2(t,h) + \cdots.
\eeq
Thus knowing $\Del \rho^{kz}(h)$ allows us to calculate 
$f(t)$ perturbatively in $\del\tilde h_3$. 
We can also expand $\Del \rho^{kz}(h)$ in $\del\tilde h_3$ and 
$\varepsilon$. This makes
the integration on the left-hand side 
of eq.\ \rf{feq5} possible order by order
in terms of elementary functions. The expansion of $\Del \rho^{kz}(h)$
is based on the following two observations. Firstly, we have for any
(not just infinitesimal) $\varepsilon$ the exact relation 
\beq{expand1}
D^{kz}(h;\a {\rm e}^\varepsilon) = D^{kz}(h;\a) + \varepsilon\, \cG (h),
\eeq
\beq{expand2}
\cG(h) = \frac{1}{i\pi}\, r(h) 
\int_{h_3}^\infty {\d h'} \, \frac{1}{(h-h') r(h')}.
\eeq
Next, $D^{kz}(h;h_3)$ and $\cG(h)$ have expansions of 
the form \rf{expanD} with respect to their $h_3$-dependence. 
Thus $\Del \rho^{kz}(h)$ can be written as
\beq{exprho}
\Del \rho^{kz} (h;\del\tilde h_3,\varepsilon) = 
(a_1\,\del\tilde h_3 +b_1\, \varepsilon)+ 
(a_2\, \del \tilde h_3 +b_2\,\varepsilon )(\del\tilde h_3)^2+ \cdots,
\eeq
where only $b_1$ is not an elementary function of $h$.
Explicitly, we have
\beq{beq}
\Del \rho^{kz}(h) =
\left(-\del\tilde h_3\,\frac{ir(h)}{2(h-h_3)}F_3(h_3)-\varepsilon\,
\cG(h,h_3)\right) +\cdots.
\eeq
Here and in the following we use the shorthand notation $F_3(h_3)$ for
the function $F_3(h_1,h_2,h_3,\a,\b)$.
The integral in \rf{feq5} can now be performed and one obtains
\beq{geq3}
f(t)=\left(C' \sqrt{t-h_3} +
\del \tilde h_3\,  \frac{r_0(h_3)}{\sqrt{t-h_3}}F_3(h_3)-
\varepsilon \right)+\cdots,
\eeq
where $C'= C+\del\tilde h_3\, F_3(h_3) + \varepsilon 
\frac{1}{i \pi} \int_{h_3}^{\infty} \frac{dh'}{r(h')}$.
Note that until this point we have not made use of any of
the boundary conditions, and therefore could treat 
$\del \tilde h_3$ and $\varepsilon$ as independent.

Finally we can insert $f(t)$ in \rf{DD1a} and \rf{DD1b} to obtain 
$D_{1,2}(h)$. To order $(\del\tilde h_3)^2$ we obtain
\bea\label{D12}
\lefteqn{D_j(h;h_1,h_2,h_3^{j};\a_j,\b) = 
D^{kz}(h;h_1,h_2,h_3;\a,\b)+}\\  \no
&&\mp \varepsilon \pm C'\left( \sqrt{h-h_3} + O(\del\tilde h_3) \right)
\pm \del\tilde h_3 \, \frac{i r_0(h_3) F_3(h_3)}{2\sqrt{h-h_3}}\\ \no
&&+\frac{(\del\tilde h_3)^2}{4}\frac{i r_0(h)}{\sqrt{h-h_3}}
\left(\frac{\partial F_3(h_3)}{\partial h_3}\pl \frac{F_3(h_3)}{2(h-h_3)}
\pl \frac{F_3(h_3)}{2(h_3-h_1)}\pl \frac{F_3(h_3)}{2(h_3-h_2)})\right)\dots
\eea
Note first that $C'\equ 0$ since the large-$h$ asymptotics 
\beq{CS1}
D_j(h) \sim i{\cal W}_j h^{1/2} -\log{\a_j h} +i
{\cal O}_j h^{-1/2}+O(1/h),
\eeq
must be satisfied for
{\it both} $D_j$'s to zeroth (and, of course, to any) 
order in $\del\tilde h_3$. 
Furthermore, note that the $\varepsilon$-dependence to this
order is simply due 
to a shift in the argument from $\log\a$ to $\log\a_j\equ
\log \a \,\pm\varepsilon$ in the terms $\log ((h-h_1)/(h^2\a_j))$
in $D_j$.


In order to give a more detailed discussion of the continuum
limit, we will also need the explicit form of the boundary conditions 
\rf{fifth}, \rf{cond1} and \rf{cond2} to second order 
$(\del\tilde h_3)^2$. From \rf{im} we get
\beq{CSim}
\oh\,\Im [D_1(h)+D_2(h)] =-i \pi \rho(h),~~~h \in I_0,
\eeq
which implies that
\beq{CS0}
\rho(h)=\rho^{kz}(h;h_1,h_2,h_3;\a,\b)+O((\del\tilde h_3)^2),
\eeq
where the explicit form of the $O((\del\tilde h_3)^2)$-corrections can be 
read off from \rf{D12}.
An expansion of the constraints ${\cal W}_j$ to second
order yields
\bea\label{CS2}
{\cal W}_j&=&W^{kz}(h_1,h_2,h_3,\a,\b)\\
&&+\frac{(\del\tilde h_3)^2}{4}
\left(\frac{\partial F_3(h_3)}{\partial h_3}
\pl \frac{F_3(h_3)}{2(h_3-h_1)}\pl \frac{F_3(h_3)}{2(h_3-h_2)})\right)
+ \dots=0.
\no\eea
In a similar way we obtain explicit expressions for ${\cal O}_j$ to 
order $(\del\tilde h_3)^2$.

These expressions can now be used to refine the relations
\rf{solu5} which govern the behaviour along an arbitrary,
non-tangential approach to the critical line. One obtains
\bea\label{Csolu2}
\del \a \sim  \del\b&\sim &\del h_2 , \nonumber \\
(\del h_3)^2 +(\del\tilde h_3)^2 &\sim & a_1\del \a+a_2\del\b \nonumber \\
\del h_3 \cdot \del\tilde h_3  &\sim& \del\tilde\a,
\eea
for some real $a_i$.
These relations can be diagonalized to give
\beq{Csolu3}
(\del h_3 \pm \del\tilde h_3)^2 \sim (\del h_3^{(i)})^2 
\sim a_1 \del\a +a_2\del\b \pm a_3 \del\tilde{\a},
\eeq
characterizing the critical behaviour related to the approach 
to the two critical surfaces discussed earlier. 

By contrast, in phase B we have only a single smooth
critical hypersurface characterized by ${\cal F}_2^{(j)}\equ 0$.
Again, moving tangentially to this surface will induce
linear variations which are all of the same order, whereas
generic, non-tangential approaches to the surface
will be characterized by
\beq{Csolu4}
\del\a\sim\del\b\sim\del\tilde\a\sim \del h_3\sim\del
\tilde h_3\sim (\del h_2)^2.
\eeq

\subsection{The scaling relations and renormalization of 3D gravity}

We now want to relate the scaling limit of the 
matrix model discussed above to 
the continuum limit of the discretized 3D quantum gravity. 
(Euclidean) quantum field theories can be defined 
as the scaling limit of suitable discretized statistical theories.
The continuum coupling constants are then  defined by a specific 
approach to a critical point of the statistical theory. Different 
approaches to the critical point
might lead to different coupling constants or even 
different continuum theories. 
In our  case we want to show that it is 
possible to approach a critical point in 
such away that the canonical scaling expected 
from a theory of 3D gravity is reproduced.

We saw above that the asymmetric model is defined on a 
three-dimensio\-nal hypersurface in the parameter space of
$(h_1,h_2,h_3^{(i)})$. Let us consider a curve
approaching a specific symmetric point $(h_1^c,h_2^c,h_3^{(i)}
\equ h_3^c)$ 
on the critical surface of the model, with curve parameter $a$
(where $a\equ 0$ corresponds to the critical point).
In terms of independent parameters, this path can be
described by $(h_2(a),h_3^{(1)}(a),h_3^{(2)}(a))$ or,
equivalently, $(\a_1(a),\a_2(a),\b (a))$. Recalling our
discussion in Sec.2 above, we are interested in obtaining
a behaviour of the form 
\bea\label{a1a2}
\log \a_i& = & \log\a_c + 
\frac{a\, c_1}{G}- a^2 Z_i-a^3 c_2 \L \\
\log\b & = & \log \b_c -\frac{a\, c_3}{G}-a^3 c_4\L\no
\eea
as $a \to 0$.
In \rf{a1a2}, $\log\a_c$ and $\log\b_c$ represent combined additive 
renormalizations of all the coupling constants, and $G$, 
$\L$ and $Z_i$ are the renormalized gravitational and
bulk and surface cosmological coupling constants.

For the purposes of conventional
quantum gravity we are primarily interested in phase A
of the model. Moreover, because of the symmetry between
in- and out-states which is related to our three-dimensional
geometric interpretation of the essentially two-dimensional
matrix model, we are only interested in continuum limits
where both of the discrete spatial boundary volumes go to
infinity, that is, where both $z_t$ and $z_{t+a}$ are
critical. Interesting critical points therefore must lie
on the critical line where the two critical surfaces intersect.

As we have argued at length in \cite{newpaper}, a 
standard critical behaviour of 3D gravity governed by the canonical 
scaling of the three-volume can only be achieved
by a tangential approach to the critical line (in that case, in
the $\beta$-$\alpha$ plane). This approach ensures that 
the terms proportional to $a/G$ in the expansions of $\a$ and
$\b$ by themselves do not determine the leading critical
behaviour, although they contribute in the combination
$\L G^3$ at higher order. The same construction carries over to 
the asymmetric case, which differs from the symmetric
situation by the appearance of the second-order terms
proportional to $Z_i$ in \rf{a1a2}. As happened there,
the tangent vector to the critical curve (in the hyperplane $\tilde
\a\equ 0$) at a given point $\a^c(\b)$ fixes the ratio
$c_1/c_3$ of the constants appearing in \rf{a1a2}.
Moreover, the leading critical behaviour will now be governed
by the boundary cosmological terms. This means in practice
that in order to determine the singular behaviour of the
free energy $F(\a_i,\b)$ under such an approach, we can simply 
ignore the order-$a$
terms in \rf{a1a2} and use the scaling relations \rf{Csolu2} 
for the remainder.

We still have to discuss the relative scaling behaviour
of $\del\tilde\a$ and $\del\a$. As we will show in the next
section, if we want the transfer matrix to reduce to
the unit operator in the limit $a\rightarrow 0$, the
singular behaviour should to leading order depend only
on $\del\a$ and not on $\del\tilde\a$. One way of realizing
this possibility would be by showing that at this order 
the coefficient in
front of the $\del\tilde\a$-term vanishes, implying the
relations
\bea\label{a1a5}
\del \a &\propto&  a^2, \\
\del\tilde\a &\propto& a^3  \no
\eea
and  consequently 
\bea\label{a1a6}
\del h_3 &\propto& a,\\
\del \tilde h_3 &\propto& a^2. \no
\eea
The corresponding scaling curve deviates
from the symmetric plane $\a_1\equ\a_2$ only by $O(a^3)$-terms. 
Note that this means that the asymmetry $\del\tilde\a$ contributes at
the same order as the cosmological term $\L$, which presumably
is a desirable property in view of the
construction of the Hamiltonian.

\section{The transfer matrix}\label{discuss}

The transfer matrix contains the information necessary to derive the 
Hamiltonian of 3D gravity. From the free energy of the asymmetric ABAB
model we can extract some information about the transfer matrix as is clear 
from formula \rf{II.1}. Let us be more precise about this (see 
\cite{ajlv} for a  detailed discussion). The free energy of the 
asymmetric ABAB matrix model involves according to \rf{II.1} a 
summation over the individual geometric states $|g\ra$ which label 
in- and out-states. However, one can use the free energy 
to extract information about the 
areas $N_{in}$ and $N_{out}$ (the number of squares in the 
in- and out-quadrangulations) 
of the in- and out-states $|g_{in}\ra$ and $|g_{out}\ra$.
We expect this quantity to capture the essential part of physical
information about the time evolution of a two-dimensional universe
(cf. e.g. \cite{carlip}).   
Let us consider the state
\beq{D1}
| N \ra = \frac{1}{\sqrt{\cN (N)}}\sum_{g_t} \delta_{N,N(g_t)} |g_t>
\eeq
where ${\cal N}(N)$ is the number of quadrangulations 
of given area  $N$. The norm of such a state is
\beq{D2}
\la N' | N \ra = \del_{N,N'}
\eeq
since states $|g_1\ra$ and $|g_2\ra$ with different quadrangulations 
are orthogonal. The number of quadrangulations constructed 
from $N$ squares grows exponentially as 
\beq{D3}
{\cal N}(N) = N^{-7/2}\; \e^{\mu_0 N} (1 + O(1/N^2)),
\eeq
where $\m_0$ is known. The sum \rf{II.1} can now be 
written as 
\beq{D4}
F(\a_1,\a_2,\b)\! =\!\!\!\! \sum_{N_t,N_{t+a}}\!\! 
\e^{-z_t N_t -z_{t+a} N_{t+a}}
\la N_{t+a}| \hT |N_t\ra \sqrt{\cN (N_t) {\cN} (N_{t+a})}.
\eeq
The exponential part of $\sqrt{{\cN}(N_t){\cN} (N_{t+a})}$ can be absorbed in 
additive renormalizations of the boundary cosmological constants 
$z_t$ and $z_{t+a}$ (i.e. additive renormalizations of $\log \a_i$,
recall \rf{II.1}). It follows that in the scaling limit, 
i.e.\ for large $N$ where we can use \rf{D3}, the Laplace transforms 
of the matrix elements $\la N_1|\hT |N_2\ra$ are  equal to the 
``7/2'' fractional derivative\footnote{There are standard ways to 
define the concept of a fractional derivative, see for instance 
\cite{fractional}.} of the free energy $F(\a_1,\a_2,\b)$,
\beq{D6}
\sum_{N_t,N_{t+a}} \e^{-z_t N_t -z_{t+a} N_{t+a}}
\la N_{t+a}| \hT |N_t\ra = 
\left(\frac{\partial}{\partial z_t}
\frac{\partial}{\partial z_{t+a}}\right)^{7/4}
F(\a_1,\a_2,\b).
\eeq 
where $-\frac{\partial}{\partial z} = \frac{\partial}{\partial \log \a}$
as is clear from \rf{II.2}.

The scaling limit, and thus the continuum physics, is determined
by the singular part of the free energy. The leading behaviour 
of this singular part when we approach a critical point as described 
in the previous section, is given by 
$F(\a_1,\a_2,\b) \propto (\del \a)^{5/2}$. It is now straightforward 
to apply \rf{D6} and one finds
\beq{D7}
\left(\frac{\partial}{\partial \log \a_1}\right)^{7/4}
\left(\frac{\partial}{\partial \log \a_2}\right)^{7/4}
F(\a_1,\a_2,\b) \approx \frac{1}{\del\a}.
\eeq
This is exactly the leading-order behaviour we expect for the transfer matrix
when $a \to 0$ from \rf{0.1}:
\beq{delta}
\la N_1| \e^{-a \hH} |N_2\ra \to \la N_1|\hat{I}|N_2\ra = \del_{N_1,N_2},
\eeq
and the Laplace transform of $\del_{N_1,N_2}$ is (for large $N$'s)
\beq{ldelta}
\frac{1}{\del\a_1+\del\a_2} = \frac{1}{a^2(Z_1+Z_2)} (1+O(a)).
\eeq
 
It would be very interesting to expand to next order in $a$ and thus 
obtain information about $\hat H$. These terms come from the 
$O(a^3)$-terms discussed in the previous section.
This will only give 
us information about the matrix elements related to the states 
of the form \rf{D1}, but as discussed in detail in \cite{ajlv}
we expect that this is the only information relevant in the continuum
limit of 3D gravity if the topology of the spatial slices is that
of a sphere.

\section*{Acknowledgement}

All authors acknowledge support by the
EU network on ``Discrete Random Geometry'', grant HPRN-CT-1999-00161.
In addition, J.A. and J.J. were supported by ``MaPhySto'',
the Center of Mathematical Physics
and Stochastics, financed by the
National Danish Research Foundation.
J.J. acknowledges  support by Polish Committee for Scientific Research
(KBN) grant 2~P03B~096~22 (2002-2004).


\begin{thebibliography}{99}



\bibitem{ajlv}J. Ambj\o rn, J. Jurkiewicz, R. Loll and G. Vernizzi:
{\it Lorentzian 3D gravity with wormholes via matrix models},
JHEP 0109 (2001) 022 [hep-th/0106082].

\bibitem{ajl1} J.\ Ambj\o rn, J.\ Jurkiewicz and R.\ Loll:
{\it A nonperturbative Lorentzian path integral for gravity},
{Phys.\ Rev.\ Lett.}\ 85 (2000) 924-927 [hep-th/0002050].

\bibitem{ajl1.5}
J.~Ambj\o rn, J.~Jurkiewicz and R.~Loll:
{\it Dynamically triangulating Lorentzian quantum gravity},
Nucl.\ Phys.\ B\ 610 (2001) 347-382 [hep-th/0105267].

\bibitem{al} J.\ Ambj\o rn and R.\ Loll:
{\it Non-perturbative Lorentzian quantum gravity, causality and 
topology change},
{Nucl.\ Phys.\ B}\ 536 (1998) 407-434 [hep-th/9805108].

\bibitem{ajl2} J. Ambj\o rn, J. Jurkiewicz and R. Loll:
{\it Nonperturbative 3-d lorentzian quantum gravity},
{Phys.\ Rev.\ D} 64 (2001) 044011 [hep-th/0011276].

\bibitem{ajl2.5}
J.~Ambj\o rn, J.~Jurkiewicz and R.~Loll:
{\it Computer simulations of 3d Lorentzian quantum gravity},
Nucl.\ Phys.\ Proc.\ Suppl.\ 94 (2001) 689-692 [hep-lat/0011055];
{\it 3d Lorentzian, dynamically triangulated quantum gravity},
Nucl.\ Phys.\ Proc.\ Suppl.\ 106 (2002) 980-982
[hep-lat/0201013].

\bibitem{kz} 
V.A.\ Kazakov and P.\ Zinn-Justin:
{\it Two matrix model with ABAB interaction.},
Nucl.\ Phys.\ B 546 (1999) 647-668 [hep-th/9808043]. 

\bibitem{ksw}
V.A.\ Kazakov, M.\ Staudacher and T.\ Wynter:
{\it Exact solution of discrete two-dimensional R$^2$ gravity},
Nucl.\ Phys.\ B\ 471, (1996) 309-333 [hep-th/9601069];
{\it Almost flat planar diagrams},
Commun.\ Math.\ Phys.\ 179 (1996) 235-256 [hep-th/9506174];
{\it Character expansion methods for matrix models of dually
weighted graphs},
Commun.\ Math.\ Phys.\ 177 (1996) 451-468 [hep-th/9502132];
P.\ Zinn-Justin:
{\it Random hermitian matrices in an external field},
Nucl.\ Phys.\ B\ {497} (1997) 725-732 [cond-mat/9703033].

\bibitem{newpaper}J. Ambj\o rn, J. Jurkiewicz and R. Loll:
{\it Renormalization of 3d quantum gravity from matrix models},
[hep-th/0307263].

\bibitem{newPZJ} P. Zinn-Justin: {\it The asymmetric ABAB matrix model}
[hep-th/0308132].

\bibitem{ackl} J.\ Ambj\o rn, J.\ Correia, C.\ Kristjansen and
R.\ Loll: {\it The relation between Euclidean and Lorentzian 2D quantum 
gravity},
{Phys.\ Lett.\ B}\ 475 (2000) 24-32 [hep-th/9912267].

\bibitem{book}N.I.\ Muskhelishvili:
{\it Singular Integral Equations}, Dover Publn. 1992.

\bibitem{fractional}S. G. Samko, A. A. Kilbas and O. I. Marichev: 
{\it Fractional integrals and derivatives: theory and applications},
Gordon \& Breach, 1987.

\bibitem{carlip}S. Carlip:
{\it Quantum gravity in 2+1 dimensions},
Cambridge Univ. Press, Cambridge, UK (1998).

\end{thebibliography}
\end{document}